\documentclass[twocolumn]{jpsj3} 
\usepackage{txfonts}

\def\Jbf{\mbox{\boldmath $J$}}

\def\Mbf{\mbox{\boldmath $M$}}
\def\Hbf{\mbox{\boldmath $H$}}

\title{Isotropic $\Gamma_6$ Ground State in Caged Compound NdRu$_2$Zn$_{20}$}

\author{
\name{Yosikazu \surname{Isikawa}\thanks{E-mail address: isikawa@sci.u-toyama.ac.jp}},
\name{Jun-ichi \surname{Ejiri}},
\name{Toshio \surname{Mizushima}}, and
\name{Tomohiko \surname{Kuwai}}
}

\inst{Graduate School of Science and Engineering, University of Toyama, Toyama 930-8555, Japan}

\abst{
The magnetic susceptibility $\chi$, the magnetization $M$, and the specific heat $C$ were measured for the caged cubic compound NdRu$_2$Zn$_{20}$ at temperatures down to 0.5 K
in the magnetic field $H$ along the three crystallographic principal axes, [100], [110], and [111].
The ferromagnetic phase transition was observed at the Curie temperature $T_{\rm C}=1.9$ K in both the temperature dependence of $\chi(T)$ and $C(T)$.
Below $T_{\rm C}$, the magnetization curves $M(H)$ show very weak magnetic anisotropy, 
and the $C(T)$ curves in the fields along the three principal axes show no magnetic anisotropy both below and above $T_{\rm C}$.
The easy direction of magnetization at 0.5 K changes from the [111] direction in 0 T to the [110] diretion in 7 T.
We analyzed these characteristic behaviors theoretically
taking into account the crystalline-electric-field energy, the magnetic exchange interaction, and the Zeeman energy.
It was found that these features are originated from the isotropic $\Gamma_6$ ground state
mixing with the magnetic $\Gamma_8^{(1)}$ excited state.
The temperature and the magnetic field dependences of $M(T,H)$ and $C(T,H)$ are quantitatively well analyzed  by 
this theoretical calculation. }

\kword{NdRu$_2$Zn$_{20}$, magnetization, specific heat, anisotropy, $\Gamma_6$ ground state, caged compound, crystalline electric field}
\begin{document}
\maketitle

The  RT$_2$Y$_{20}$-type compounds have recently attracted much attention because of their variety of physical properties,
where R is a rare-earth atom, T is a transition-metal atom, and Y is Zn or Al. For example,
CeRu$_2$Zn$_{20}$ is one of the dense Kondo compounds with an enhanced electronic-specific-heat coefficient.\cite{isi1}
PrRh$_2$Zn$_{20}$ shows a quadrupolar order at an extremely low temperature.\cite{oni1} 
PrIr$_2$Zn$_{20}$, PrRh$_2$Zn$_{20}$, LaIr$_2$Zn$_{20}$ and LaRu$_2$Zn$_{20}$ are superconductors.\cite{oni1,naga,oni2}
YbT$_2$Zn$_{20}$ (T = Fe, Co, Ru, Rh, Os, Ir) are heavy-electron compounds with large local-moment degeneracy.\cite{tori}
DyFe$_2$Zn$_{20}$ is a ferromagnetic compound with a strong coupling between the Dy and Fe ions.\cite{isi2,tamu}
These compounds are crystallized in the cubic CeCr$_2$Al$_{20}$-type structure (space group is Fd$\bar 3$m).\cite{nasc}
R element seizes in the cubic-symmetry site (the point symmetry is T$_{\rm h}$).
Characteristic features in this type of compounds are (1) the weak magnetic coupling due to the dilute content of R elements,
and (2) the weak magnetic anisotropy due to the highly symmetrical environment around R ions.\cite{nasc} 

PrRu$_2$Zn$_{20}$ remains in a normal state down to 0.04 K,
showing no magnetic order nor quadrupolar order but a structural transition at 138 K, according to Onimaru et al\cite{oni1}.
The energy  scheme due to the crystalline electric field (CEF) is speculated to be a singlet ground state and a singlet excited state situated at 8 K
above the ground state based on the low-temperature specific-heat experiment.\cite{oni1}
Very recently, Iwasa et al.\cite{iwas} determined the energy scheme and the CEF parameters $x$ and $W$ of PrRu$_2$Zn$_{20}$ 
by the inelastic neutron scattering experiment. Their ground state and excited state are $\Gamma_3$(0 K)-$\Gamma_5$(36.7 K),
which is contradictory to the energy scheme inferred from the specific-heat experiment.

We have studied the magnetic and the thermal properties of NdRu$_2$Zn$_{20}$ 
to observe the competition between the exchange interaction and the CEF effect.
In this Letter, we present the experimental results of NdRu$_2$Zn$_{20}$ at temperatures down to 0.5 K.
NdRu$_2$Zn$_{20}$ is a ferromagnet with the Curie temperature $T_{\rm C}=1.9$ K.
Present experimental results could be analyzed numerically very well by taking into account the exchange interaction between the Nd ions, the CEF effect, and the Zeeman energy.
The CEF parameter of NdRu$_2$Zn$_{20}$ was derived by a sophisticated method based on the CEF parameters of PrRu$_2$Zn$_{20}$.\cite{iwas}
Theoretical calculations reveal that the isotropic properties of NdRu$_2$Zn$_{20}$ 
 are caused by the isotropic $\Gamma_6$ ground state
and that the weak anisotropic properties are originated from the mixing effect with the magnetic $\Gamma_8$ excited state.

Single crystals of NdRu$_2$Zn$_{20}$ were grown by the Zn-self-flux method
which was the same as that described previously.\cite{isi1,isi2}
Crystal structure of the cubic CeCr$_2$Al$_{20}$ type was confirmed by the X-ray powder diffraction pattern. 
There was no trace of impurity phases. 
The lattice parameter $a$ was obtained to be 14.321\AA~from the X-ray diffraction pattern, 
which is in good agreement with the value of a literature\cite{nasc}.
Crystal orientation was determined by Laue pictures. 
The same single crystal was used in both the experiments of the magnetization and the specific heat.
The sample mass is 1.31 mg.
The magnetization $M$ and the magnetic susceptibility $\chi$ were
measured at temperatures down to 0.5 K by using 
a magnetic properties measurement system (MPMS, Quantum design Ltd.).
The specific heat $C$ was measured at temperatures down to 0.5 K by using a physical properties measurement system 
(PPMS, Quantum design Ltd.).

Figure 1 shows the magnetization curves $M(H)$ in the magnetic fields $H$
along the three crystallographic principal axes [100], [110], and [111]  up to 7 T at various temperatures. 
As seen in this figure, NdRu$_2$Zn$_{20}$ is a ferromagnet.
$T_{\rm C}$ is 1.9 K, which will be accurately determined from the $C(T)$ curve later.
Crystallographic anisotropy is extremely small even at temperatures below $T_{\rm C}$. 
At 0.5 K in $H$ above 1 T, both $M$'s along the [111] and the [110] directions are slightly larger than that of $M$ along the [100] direction, 
indicating that the hard direction of magnetization is the [100] direction. 
Furthermore, $M$ along the [110] direction is more slightly larger than that along the [111] direction, 
suggesting that the easy direction of magnetization is the [110] direction in $H$ above 1 T.
The ferromagnetic moment at 0.5 K, extrapolated smoothly to 0 T, is approximately 1.46 $\mu_{\rm B}$/Nd, 
which is less than that of the free Nd ion 3.28 $\mu_{\rm B}$.
This reduced moment is caused by the CEF effect, as will be discussed later.

The inset in Fig. 1 shows the $M(H)$ curves at 0.5 K in low $H$ below 0.2 T.
The magnetic anisotropy was apparently observed; the easy direction of magnetization is the [111] direction, and the hard direction of magnetization is the [100] direction. 
As regards the magnetization in low $H$, we have to pay much attention to the effects of hysteresis, demagnetizing field, 
domain-wall motion, domain rotation, etc.
We confirmed that there were no hysteresis in $M(H)$ curves within the experimental errors.
The demagnetizing field is not corrected in this figure, i. e., $H$ is an external magnetic field, 
because the shape of the sample is not regular rectangular nor spherical. 
The demagnetizing field is roughly estimated to be 0.015 T in case of $M=1.2~\mu_{\rm B}$/Nd if we assume that the sample shape is spherical.
Neglecting these effect, 
the magnetizations extrapolated smoothly to 0 T, $M_0$'s, are estimated to be approximately 
1.2, 1.0, and 0.8 $\mu_{\rm B}$/Nd for the [111], the [110], and the [100] directions, respectively.
The ratio of these values is roughly equal to $1:\sqrt{2/3}:\sqrt{1/3}$,
indicating that 
(1) the easy direction of magnetization is the [111] direction, and 
(2) $M_0$'s along the [110] and the [100] directions are the projection components of $M_0$ along the [111] direction.
The easy direction of magnetization [111] below 0.2 T is contradictory to the easy direction [110] in 7 T.

We measured the temperature dependence of the magnetization $M(T)$ in low $H$ along the three principal axes.
As shown in Fig. 2, 
$M(T)$ in 0.1 T shows a gradual increase in the vicinity of $T_{\rm C}$ with decreasing temperature.
$M(T)$ in lower $H$, 0.01 T, shows a steep increase at $T_{\rm C}$.
The inset in Fig. 2 shows the temperature dependence of the inverse magnetic susceptibility $1/\chi(T)$,
which obeys the Curie-Weiss law well at the temperature range above 50 K. 
The paramagnetic Curie temperature ${\it \Theta}_{\rm p}$ is $2 \pm 2$ K
and the effective Bohr magneton $\mu_{\rm eff}$ is 3.62 $\pm 0.01$ $\mu_{\rm B}/$ion. 
The value of $\mu_{\rm eff}$ is in good agreement with that of the free Nd$^{+3}$ ion 3.62 $\mu_{\rm B}$.
\begin{figure}[h]
\begin{center}
\includegraphics[width=65mm]{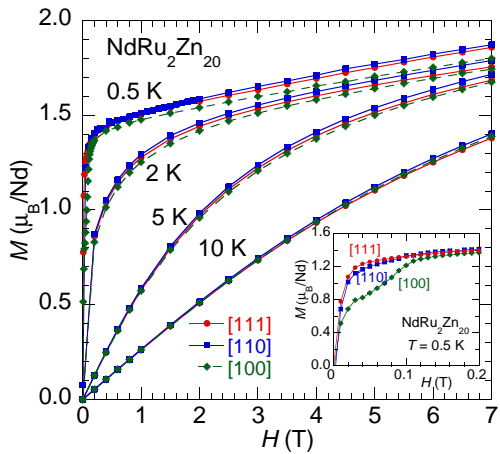}
\caption{\label{label}(Color online) Magnetization curves of NdRu$_2$Zn$_{20}$ in $H$ along the three crystallographic principal axes.
Inset shows the magnetization curves in $H$ below 0.2 T.}
\end{center}
\end{figure}
\begin{figure}[h]
\begin{center}
\includegraphics[width=65mm]{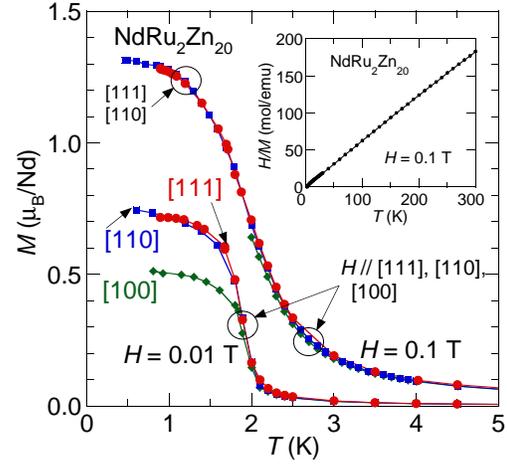}
\caption{\label{label}(Color online) Temperature dependence of the magnetization of NdRu$_2$Zn$_{20}$ in $H=0.1$ and 0.01 T along the three crystallographic principal axes. 
The inset is the temperature dependence of the inverse magnetic susceptibility in $H=0.1$ T along the [111] direction.}
\end{center}
\end{figure}


Figure 3 shows the temperature dependence of the specific heat $C(T)$ of NdRu$_2$Zn$_{20}$ in various $H$.
The $\lambda$-type anomaly in 0 T was observed at $T_{\rm C}=1.9$ K, which is in good agreement with the one speculated in the $M(T)$ curves. 
With applying $H$, this sharp anomaly changes to a round peak, and the peak position shifts to the higher temperatures.
The entropy $S_{\rm 4f}$ in $H=0$ is estimated to be approximately $R$ ln 2 near $T_{\rm C}$,
indicating that the ground state of Nd ions is doublet.
It is worth noting that no anisotropy of $C(T)$ is observed at $T$ between 0.5 and 3 K when $H$ is applied along the three principal axes, [100], [110] and [111], as shown in Figs. 3 (a), (b), and (c).

\begin{figure}[h]
\begin{center}
\includegraphics[width=85mm]{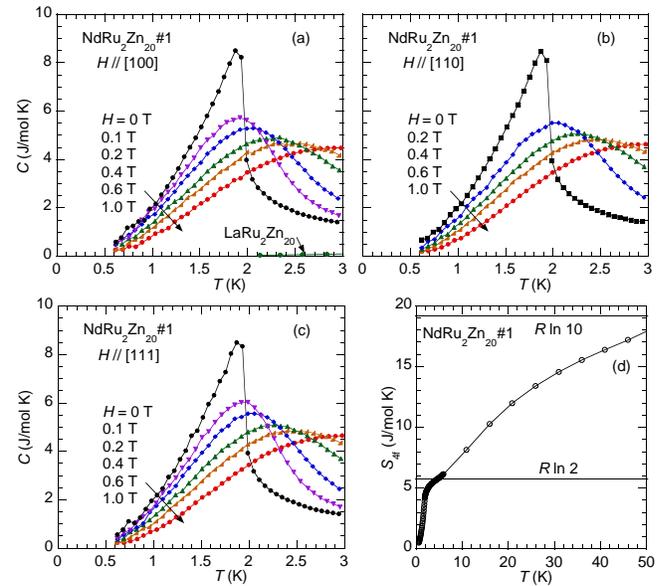}
\caption{\label{label}(Color online) Temperature dependence of the specific heat of NdRu$_2$Zn$_{20}$
in $H$ along the [100] direction (a), the [110] direction (b), and the [111] direction (c). The temperature dependence of the specific heat of LaRu$_2$Zn$_{20}$ is also shown in (a). (d) shows the temperature dependence of the magnetic part of the entropy of NdRu$_2$Zn$_{20}$ in $H=0$ T.}
\end{center}
\end{figure}

We define the following Hamiltonian to analyze the magnetic and thermal properties of NdRu$_2$Zn$_{20}$,\cite{isi2}
\begin{eqnarray}
{\cal H}= {\cal H}_{\rm CEF}+ {\cal H}_{\rm exch} + {\cal H}_{\rm Z} ,
\end{eqnarray}
where ${\cal H}_{\rm CEF}$ is the CEF Hamiltonian, ${\cal H}_{\rm Z}$ the Zeeman Hamiltonian, 
and ${\cal H}_{\rm exch}$ the exchange Hamiltonian between the two Nd atoms. 
The exchange interaction is treated in the frame of the molecular-field approximation.
${\cal H}_{\rm CEF}$ is written as\cite{LLW}
\begin{eqnarray}
{\cal H}_{\rm CEF} = W\left(x\frac{O_4}{F(4)} +(1-|x|)\frac{O_6}{F(6)}\right) ,
\end{eqnarray}
where $O_4=O_4^0+5~O_4^4$ and  $O_6=O_6^0-21~O_6^4$.
Here the z axis is the [001] direction.
${\cal H}_{\rm Z}$ and  ${\cal H}_{\rm exch}$ are expressed as
\begin{eqnarray}
{\cal H}_{\rm Z} &=&  -\Mbf_{\rm R}\Hbf_{\rm ext} ,\\
{\cal H}_{\rm exch} &=&  -\Mbf_{\rm R}\Hbf_{\rm mol} ,
\end{eqnarray}
where $\Mbf_{\rm R}$ is the magnetization of Nd atoms, $\Hbf_{\rm ext}$ the external magnetic field, and
$\Hbf_{\rm mol}$ the molecular field at Nd atom caused by surrounding Nd atoms.
These parameters are defined as follows:
$\Mbf_{\rm R} = -g_J\mu _{\rm B}\Jbf$ and  $\Hbf_{\rm mol} = n_{\rm RR}\langle \Mbf_{\rm R}\rangle$ where
$\Jbf$ is the total angular momentum, and $n_{\rm RR}$ the molecular-field parameter between Nd atoms.
For simplicity, the $n_{\rm RR}$ is chosen as a scalar.
Note that the physical quantities of $\Mbf_{\rm R}$, $\Jbf$, $\Hbf_{\rm mol}$, and $\Hbf_{\rm ext}$ are three-dimensional vectors. 
The angular momentum $\Jbf$ is calculated by
\begin{eqnarray}
\langle \Jbf\rangle= \frac{{\rm Tr}~ \Jbf \exp (-\beta {\cal H})}{{\rm Tr}~ \exp (-\beta {\cal H})} ,
\end{eqnarray}
where $\beta=1/k_{\rm B}T$ and $k_{\rm B}$ denotes the Boltzmann constant. 
The Eq. 5 is numerically solved by an usual procedure of iteration. 
The specific heat $C$ is calculated by
\begin{eqnarray}
C= N_{\rm A}\frac{\partial }{\partial T}\left(\langle {\cal H}\rangle -\frac{1}{2}\langle {\cal H}_{\rm exch}\rangle \right) ,
\end{eqnarray}
where the second term in the parenthesis is a correction term to correct the double-counting of the exchange energy 
between the Nd atoms.
The $C$ is numerically calculated using the value of $\langle \Jbf\rangle$ obtained in Eq. 5.

We have three fitting parameters, i. e., $W$, $x$, and $n_{\rm RR}$ 
when we numerically calculate $M(T,H)$ and $C(T,H)$ using the above Hamiltonian,
where $H$ is the amount of $\Hbf_{\rm ext}$. 
However we can reduce the number of these fitting parameters by utilizing the energy scheme of PrRu$_2$Zn$_{20}$ determined recently by Iwasa et al.\cite{iwas}
They determined the CEF parameters of $W$ and $x$ for  PrRu$_2$Zn$_{20}$ by the inelastic neutron scattering experiment.
Two values, $W$ and $x$,  for NdRu$_2$Zn$_{20}$ are, of course, different from those of PrRu$_2$Zn$_{20}$.
However, if the following equation is employed as a CEF Hamiltonian,\cite{hutc} 
\begin{eqnarray}
{\cal H}_{\rm CEF} = A_4^0\langle r^4\rangle\beta_J  O_4 + A_6^0\langle r^6\rangle\gamma_J O_6~,
\end{eqnarray}
the CEF parameters for NdRu$_2$Zn$_{20}$ can be estimated. 
Here, two parameters $\beta_J$ and $\gamma_J$ are listed in the Hutchings's paper.\cite{hutc}
$A_4^0$ and $A_6^0$ are the parameters determined 
by the electrical potential surrounding the origin of the Nd or the Pr atom. 
Thus, the values of $A_4^0$ and $A_6^0$ for NdRu$_2$Zn$_{20}$ can be supposed to be approximately the same as those for PrRu$_2$Zn$_{20}$ especially 
because the Nd atom is the next atom of the Pr atom in the series of the rare-earth elements. 
We calculated the values of $A_4^0$ and $A_6^0$ using $W$ and $x$ of PrRu$_2$Zn$_{20}$\cite{iwas}, and then
we derived the values of $W$ and $x$ of NdRu$_2$Zn$_{20}$ using these $A_4^0$ and $A_6^0$.
The obtained $W$ and $x$ for NdRu$_2$Zn$_{20}$ are summarized in Table I together with 
the parameters used in this calculation.\cite{LLW,hutc,free}
Consequently, the fitting parameter is only $n_{\rm RR}$.
The $n_{\rm RR}$ is adjusted until the calculated $T_{\rm C}$ approaches 1.9 K.
The $n_{\rm RR}$ finally obtained is 0.51 [T/$\mu_{\rm B}$].

\begin{table}[h]
\caption{CEF parameters for NdRu$_2$Zn$_{20}$ (Nd) and  PrRu$_2$Zn$_{20}$ (Pr).\cite{iwas} 
The unit of  $W$,  $A_4^0\langle r^4\rangle$, and  $A_6^0\langle r^6\rangle$ is K.
The unit of  $\langle r^4\rangle$ and $\langle r^6\rangle$ is atomic unit.
The values of $F(4)$, $F(6)$, $\beta$, $\gamma$, $\langle r^4\rangle$, and $\langle r^6\rangle$ are cited from references.\cite{LLW,hutc,free} }
\label{table:1}
\begin{tabular}{cccccccccccc}  \hline
   & W & x &  $A_4^0\langle r^4\rangle$ & $A_6^0\langle r^6\rangle$ &  $A_4^0$ & $A_6^0$  \\    \hline
Pr &-0.626  &   0.02   &  0.2840 & -7.983 & 0.08263 & -0.4163 \\ 
Nd & 0.603  & -0.007 & 0.2405 & -6.259  & 0.08263 & -0.4163 \\ \hline 
         & $F(4)$ &  $F(6)$ & $\beta\times 10^{4}$ & $\gamma\times 10^{6}$ &$\langle r^4\rangle$&$\langle r^6\rangle$ \\  \hline 
Pr & 60 & 1260 & -7.3462 &  60.994 & 3.437 &19.17 \\ 
Nd & 60 & 2520 & -2.9111 & -37.988 & 2.910 &15.03 \\ \hline
\end{tabular}
\end{table}

Figure 4 shows the calculated $M(H)$ curves using these three parameters in $H$ along the three principal axes at 0.5, 2, 5, and 10 K.
The inset shows the calculated $M(H)$ curves in the low $H$ below 0.05 T. 
The calculated curves of $M(H)$ are excellently in good agreement with the experimental $M(H)$ (Fig. 1)
in $H$ along the three principal axes between 0 and 7 T at 0.5, 2, 5, and 10 K.
The characteristic behaviors of the experimental $M(H)$ curves are almost isotropic except the following two points;
a weak anisotropy of $M(H)$ in 7 T and an anisotropy of $M(H)$ in low $H$ below 0.05 T between the three principal axes.
These anisotropic behaviors of $M(H)$ are reproduced satisfactorily by the present calculation,
indicating that the easy and the hard directions of magnetization 
are the [111] and the [100] directions in $H=0$, respectively. 
The easy direction of magnetization changes from the [111] direction in 0 T to the [110] direction in 7 T, 
which is also reproduced by the present calculation.
It is found from the theoretical calculation 
that the isotropic magnetic behaviors are originated from the isotropic $\Gamma_6$ ground state, 
and that the weak magnetic anisotropy is caused by the magnetic $\Gamma_8^{(1)}$ excited state.
\begin{figure}[h]
\begin{center}
\includegraphics[width=65mm]{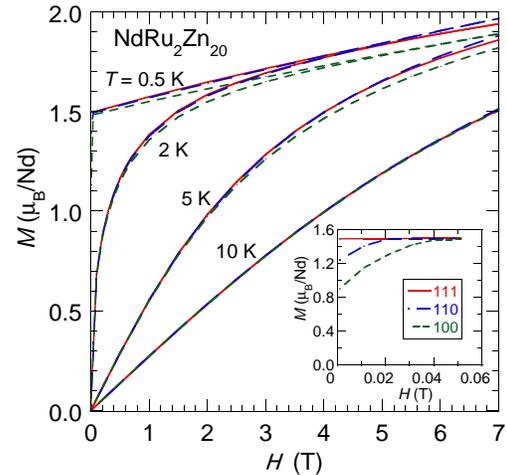}
\caption{\label{label}(Color online) Calculated magnetization curves of NdRu$_2$Zn$_{20}$ in $H$ along the [111] direction (red solid line), the [110] direction (blue long-dashed line) and the [100] direction (green dashed line). The inset shows the calculated magnetization curves at 0.5 K in $H$ below 0.05 T. }
\end{center}
\end{figure}

Next, we show the calculated $M(T)$ curves in Fig. 5 in $H=0.1$ and 0.01 T.
$M(T)$ curves in $H = 0.1$ T do not depend on the three principal axes, i. e., 
$M(T)$ is almost isotropic. The change of $M(T)$ at $T_{\rm C}$ is moderate.
In order to clarify $T_{\rm C}$, we calculated $M(T)$ in $H=0.01$ T.
As seen in the figure, 
below $T_{\rm C}$, we notice two things;
one is magnetic anisotropy, that is, 
this anisotropic $M(T)$ in low $H$ is qualitatively in good agreement with the experimental $M(T)$ (Fig. 2).
Another is the quantitative disagreement with the experimental $M(T)$.
This disagreement is due to the domain-wall motion, the domain rotation, etc., as already mentioned.
\begin{figure}[h]
\begin{center}
\includegraphics[width=55mm]{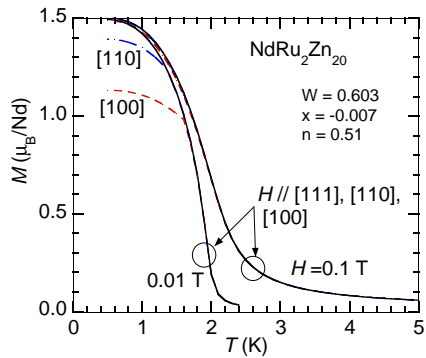}
\caption{\label{label}(Color online) Temperature dependence of the calculated magnetization of NdRu$_2$Zn$_{20}$
in $H$ along the three principal axes: [111], [110] and [100].}
\end{center}
\end{figure}

Figure 6 shows the calculated $C(T)$ curves of NdRu$_2$Zn$_{20}$ in $H$ along the [111] direction.
With increasing $H$, the sharp peak due to the $T_{\rm C}$ becomes a round peak, 
and the peak temperature moves to the higher temperatures.  
Overall behaviors of $C(T)$ are in agreement with the experimental ones (Fig. 3). 
In $H=0$, however, the calculated curve is not excellently in agreement with the experimental $C(T)$ curve, 
especially, in the higher $T$ region than $T_{\rm C}$.
In this $T$ region, the effect of the magnetic fluctuation $\langle\Jbf_{\rm i}\Jbf_{\rm j}\rangle$ is important.
However, we did not take it into account in this calculation because of the molecular-field approximation.
Except this $T$ region, 
the calculated $C(T)$ curves in $H$ are quantitatively in good agreement with the experimental $C(T)$ in the respective $H$.
The calculated $C(T)$ curves in $H$ along the [110] and the [100] directions are almost the same as the $C(T)$ curves in $H$ along the [111] direction, 
so, only calculated $C(T)$ curve in 1.0 T along the [100] direction is added in Fig. 6 by the dashed line.

\begin{figure}[t]
\begin{center}
\includegraphics[width=55mm]{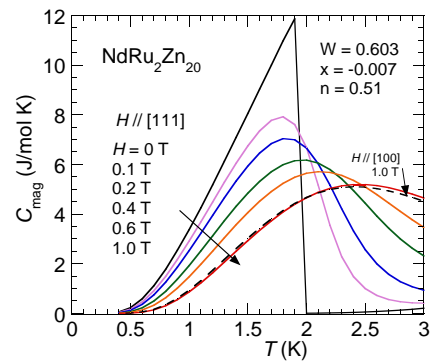}
\caption{\label{label}(Color online) Temperature dependence of the calculated specific heat of NdRu$_2$Zn$_{20}$ in $H$ along the [111] direction (solid lines). 
The dashed line denotes the temperature dependence of the calculated specific heat in $H=1.0$ T along the [100] direction.}
\end{center}
\end{figure}

We consider the calculated CEF energy scheme.
The ground state of CEF is doublet $\Gamma_6$; the wave vector is 
$\Psi(\Gamma_6)= \frac{1}{\sqrt{24}} (3|\pm \frac{9}{2}\rangle + \sqrt{14}|\pm \frac{1}{2}\rangle+|\mp \frac{7}{2}\rangle)$.
The first and the second excited states are quartets $\Gamma_8^{(1)}$ and  $\Gamma_8^{(2)}$
 situated at 26.1 K and 70.1 K above the ground state, respectively.
The expectation value of $\langle \Gamma_6| J_z|\Gamma_6 \rangle$ is $\pm$1.833.
Thus, the magnetic moment $g_J\langle J_z \rangle \mu_{\rm B}$ in the ground state is 1.33 $\mu_{\rm B}$.
Note that this magnetic moment is isotropic in the ground state, not depending on the directions of the crystal axes. 
These theoretical natures in the ground state (isotropic, and 1.33 $\mu_{\rm B}$) are slightly discrepant 
from the experimental results ($M$ is slightly anisotropic, and 1.46 $\mu_{\rm B}$ in 0 T).
These discrepancies are caused by the mixing effect, that is, the ground state is mixing with the $\Gamma_8^{(1)}$ excited state  through the exchange interaction.
Furthermore, the experimental result, that the easy direction of magnetization at 0.5 K changes from the [111] direction in 0 T to the [110] direction in 7 T, is also originated from the mixing effect with $\Gamma_8^{(1)}$ 
through both the exchange interaction and the Zeeman effect.

Finally, we have a brief discussion on why the CEF parameters in PrRu$_2$Zn$_{20}$ do not reproduce $M(T)$ and $C(T)$ very well\cite{iwas}, 
whereas they do for NdRu$_2$Zn$_{20}$. 
One possible reason is inferred as follows. The Pr ion has an integer $J$, whereas the Nd ion has a half integer $J$. 
In the case of a doublet ground state, the energy scheme of the Pr ion is easily modified by some small disturbance 
but the energy scheme of the Nd ion is not affected by a small disturbance. 

We summarize the present study.
The magnetic susceptibility $\chi$, the magnetization $M$, and the specific heat $C$ were measured for the caged cubic compound NdRu$_2$Zn$_{20}$ at temperatures down to 0.5 K
in $H$ along the three principal axes.
The ferromagnetic phase transition  was observed at $T_{\rm C}=1.9$ K in both $\chi(T)$ and $C(T)$.
Below $T_{\rm C}$, the magnetization curves $M(H)$ show a weak magnetic anisotropy. 
The easy direction of magnetization at 0.5 K is the [111] direction in 0 T, and it changes to the [110] direction in 7 T. 
The experimental results including these anisotropic properties can be analyzed numerically very well by taking into account the exchange interaction between Nd ions, the CEF effect, and the Zeeman effect.
The CEF parameter of NdRu$_2$Zn$_{20}$ was derived by a sophisticated method based on the CEF parameters of PrRu$_2$Zn$_{20}$.\cite{iwas}
Theoretical calculations reveal that the physical properties of NdRu$_2$Zn$_{20}$ are caused by the isotopic $\Gamma_6$ ground state,
and that the weak anisotropic properties are originated from the mixing effect with 
the magnetic $\Gamma_8^{(1)}$ excited state situated at 26.1 K above the $\Gamma_6$ ground state.


\vfill

\end{document}